\documentclass[numreferences]{kluwer}    % Specifies the document style.

\newdisplay{guess}{Conjecture}

\usepackage{graphicx}
\begin{document}

\begin{article}

%%%%%%%%%%%%%%%%%%%%%%%%%%%%%%%%%%%%%%%%%%%%%%%%%%%%%%%%%%%%%%%%%%%
% 
%  Our own definitions:
% 
%%%%%%%%%%%%%%%%%%%%%%%%%%%%%%%%%%%%%%%%%%%%%%%%%%%%%%%%%%%%%%%%%%% 
 
\newcommand{\be}{\begin{equation}} 
\newcommand{\ee}{\end{equation}} 
\newcommand{\mb}[1]{\mbox{\boldmath $#1$}} 
 
%%%%%%%%%%%%%%%%%%%%%%%%%%%%%%%%%%%%%%%%%%%%%%%%%%%%%%%%%%%%%%%%%%% 

\begin{opening} 

\title{Varying Fundamental Constants from a String-inspired Brane World Model} 
 
\author{Cristiano \surname{Germani}} 
 
\author{Carlos F. \surname{Sopuerta}}  
 
\institute{Institute of Cosmology and Gravitation, University of Portsmouth, 
Portsmouth PO1 2EG, Britain} 
 
\date{\today} 
 
\begin{abstract} 
We report results on the construction of cosmological braneworld models 
in the context of the Einstein-Gauss-Bonnet gravity, which include 
the leading correction to the Einstein-Hilbert action suggested 
by superstring theory.  We obtain and study the equations governing the 
dynamics of the standard cosmological models.  We find that they can be 
written in the same form as in the case of the Randall-Sundrum model but 
with time-varying four-dimensional gravitational and cosmological constants. 
Finally, we discuss the cosmological evolution predicted by these models   
and their compatibility with observational data.                            
\end{abstract}
 

\end{opening}

%%%%%%%%%%%%%%%%%%%%%%%%%%%%%%%%%%%%%%%%%%%%%%%%%%%%%%%%%%%%%%%%%%%%%% 
% 
%     SECTION I.    INTRODUCTION 
% 
%%%%%%%%%%%%%%%%%%%%%%%%%%%%%%%%%%%%%%%%%%%%%%%%%%%%%%%%%%%%%%%%%%%%%% 
 
\section{Introduction} 
Many of the cosmological scenarios nowadays under study have been                
largely originated from theoretical advances in high-energy physics. 
Of special interest is the study of the new scenarios motivated 
by developments in string and M theories, where the spacetime has non-compact 
extra-dimensions.  One of the most studied models has been proposed 
by Randall and Sundrum (RS)~\cite{RaSu}. In the RS model all matter 
and gauge fields, with the exception of gravity, are confined in a 
3-brane embedded in a five-dimensional (5D) spacetime, the {\em bulk}, 
with a $\dZ_2$ symmetry (with respect to the 3-brane).  It has been shown 
that the zero-mode of the Kaluza-Klein dimensional reduction is
localized around the 3-brane, reproducing Newtonian gravity in the 
weak field approximation.  However, the RS model does not include any 
higher-curvature correction 
predicted by string theories.  Of particular importance 
is the Gauss-Bonnet term, which is the quadratic correction allowed 
in order to have a ghost-free action~\cite{Zw,BoDe}. 
Moreover, in 5D spacetimes the most general Lagrangian producing second-order 
field equations is the combination of the Einstein-Hilbert and Gauss-Bonnet 
terms~\cite{Lo}. Following the motivation provided by these facts we have 
studied how the dynamics of the standard cosmological models, 
the Friedmann-Lema\^{\i}tre-Robertson-Walker (FLRW) models, 
is modified by the introduction of the Gauss-Bonnet term~\cite{GeSo}.

%%%%%%%%%%%%%%%%%%%%%%%%%%%%%%%%%%%%%%%%%%%%%%%%%%%%%%%%%%%%%%%%%%%%%% 
% 
%     SECTION II. 
% 
%%%%%%%%%%%%%%%%%%%%%%%%%%%%%%%%%%%%%%%%%%%%%%%%%%%%%%%%%%%%%%%%%%%%%% 
 
\section{Basic ingredients of the model} 
The field equations that one obtains from the modification of
the RS model by the Gauss-Bonnet term are (see~\cite{GeSo} for 
details):
\begin{eqnarray}\label{field} 
G_{AB}+\Lambda g_{AB}+\alpha H_{AB} = \kappa^2\left[-\lambda 
\tilde{g}_{AB}+T_{AB}\right]\delta(w)~~~(\Lambda<0)\,, 
\label{fieldeq} 
\end{eqnarray} 
where %: $H_{AB}$ is the second-order Lovelock tensor~\cite{Lo},                    
\begin{eqnarray}
H_{AB} & = & RR_{AB}-2R_A{}^CR_{BC}-2R^{CD}R_{ACBD}+R_A{}^{CDE}R_{BCDE}
\nonumber \\
& & -\textstyle{1\over4}g_{AB}(R^2-4R^{CD}R_{CD}+R^{CDEF}R_{CDEF})
\,, \nonumber
\end{eqnarray}
$\tilde{g}_{AB}$ is the 3-brane metric, $T_{AB}$ is the
energy-momentum tensor of the matter confined on the 3-brane
($w=0$) ($T_{AB}n^B=0=\tilde{g}_{AB}n^B$), and $\lambda$ is a constant
that coincides with the brane tension in the limit
$T_{AB}=0=\alpha$.   According to studies in string theory~\cite{BoDe},
the fundamental constant $\alpha$ should be positive.  To study how the
FLRW models behave in our theory we assume: (i) $T_{AB}$ is of the 
perfect-fluid type               
\begin{eqnarray}
T_{AB} = (\rho+p)u_A u_B + p\tilde{g}_{AB} 
\,,\nonumber
\end{eqnarray} 
with $u^A$, $\rho$, and $p$, being the fluid velocity ($u^A 
u_A=-1$), energy density and pressure respectively.  (ii) The                       
5D line element has the following form (introduced in~\cite{Fr}):             
\begin{eqnarray} 
ds^2 = -n^2(t,y) dt^2 + a^2(t,y)h_{ij}(k,x^l)dx^i dx^j + b^2(t,y) dy^2\,. 
\label{line} 
\end{eqnarray} 
Here $y$ is coordinate in the fifth dimension and $h_{ij}$ is a 
three-dimensional maximally symmetric metric for the surfaces 
$\{t={\rm const.},y={\rm const.}\}$, whose spatial curvature is 
parametrized by $k=-1,0,1\,.$   Then, every spacelike hypersurface                
$y={\rm const.}$ has a FLRW metric.                                               
 
The solutions of~(\ref{fieldeq},\ref{line}) in the bulk for which a               
limit in Einstein gravity exists are given by (see~\cite{GeSo} for                
details)                                                                          
\begin{eqnarray}                                                                  
\textstyle{\dot{a}\over a}\textstyle{n'\over n} +                                 
\textstyle{a'\over a}\textstyle{\dot{b}\over b} -                                 
\textstyle{\dot{a}'\over a} = 0 \,, ~                                             
\Phi + \alpha\Phi^2 = \textstyle{\Lambda\over 6}                                 
+ C(\textstyle{a_i\over a})^4\,,~                                                 
\Phi = \frac{\dot{a}^2}{a^2n^2} -                                                 
\frac{a'^2}{a^2b^2}+\frac{k}{a^2} \,.\nonumber                                    
\end{eqnarray}                                                                    
where $C$ is an integration constant and $a_i$ is a constant that later           
we will use as the initial scale factor on the 3-brane, which will                
represent the freedom in the choice of the initial time.                          
 
Given the bulk spacetime, there are two important ingredients in the              
geometrical construction of a braneworld model: the embedding of the  
3-brane and the implementation of the $\dZ_2$ symmetry. 
This can be done by following the same procedure used in the matching 
of two spacetimes in General Relativity, where two spacetimes with 
boundary are glued by using a one-to-one identification of the points
in the boundaries.  Now, the two spacetimes will be described by the 
metric~(\ref{line}), one with $y\leq 0$, say $V^-$, and the other one
with $y\geq 0$, say $V^+$.  For both spacetimes, the boundary considered 
is the hypersurface $y=0$, where the 3-brane will be located.  Now, 
instead of identifying only the boundaries of $V^+$ and $V^-$ we 
can identify all the points of $V^+$ with those of $V^-$ in 
the following way: $(x^\alpha,y)\leftrightarrow (x^\alpha,-y)$. 
With this we establish how these spacetimes have to be matched and, 
at the same time, we implement the $\dZ_2$ symmetry. 

With regard to the 3-brane, the metric that it inherits from $V^+$ 
and $V^-$ is the same, as expected.  However, the extrinsic                        
curvatures differ by a sign, due to the $\dZ_2$ symmetry, and 
hence the normal derivative of the braneworld metric has 
a jump across the 3-brane.  This jump can be determined in terms of               
energy-momentum distribution on the 3-brane [see the right-hand side of 
Eq.~(\ref{fieldeq})].  To that end, one has to use find the junction              
conditions corresponding to our theory, which are different from those 
in General Relativity.  This requires to formulate the field equations            
in the sense of distributions and to take special care of some terms              
coming from the Gauss-Bonnet term (details will be given in~\cite{GeSo2}).       
Following this procedure for the geometries we have considered~(\ref{line}),      
and considering only the solution with a proper limit $\alpha\rightarrow 0$,       
we arrive at the following modified Friedmann equation for the Hubble 
function\footnote{The subscript ``$o$'' denotes the value of the quantity on 
the 3-brane.} $H=\dot a_o/(n_oa_o)$: 
\begin{eqnarray} 
H^2 = \frac{1}{3}\tilde{\kappa}^2_\ast \rho\left(1+\frac{\rho}{2\lambda} 
\right)-\frac{k}{a^2_o}+\frac{1}{3}\tilde{\Lambda}_\ast\,. 
\label{gfeq} 
\end{eqnarray}
Here, the 4D gravitational coupling and cosmological {\em constants},             
$\tilde{\kappa}_\ast$ and $\tilde{\Lambda}_\ast$ respectively, are time           
dependent.  Actually, they change in time as functions only of                    
the scale factor $a_o$.   Their explicit form is 
\begin{eqnarray} 
\tilde{\kappa}^2_\ast & = & \frac{\tilde{\kappa}^2} 
{1+\varepsilon(\frac{a_i}{a_o})^4}\,,~~ 
\tilde{\kappa}^2 \equiv \frac{\lambda\kappa^4}{6\chi^2}\,,~~ 
\varepsilon \equiv \frac{4\alpha C}{\chi^2}\,, ~~ 
\chi\equiv \sqrt{1+\textstyle{2\over3}\alpha\Lambda}  \nonumber \\ 
\tilde{\Lambda}_\ast & = & \frac{\lambda}{2}\frac{\tilde{\kappa}^2} 
{1+\varepsilon(\frac{a_i}{a_o})^4}-\frac{3}{2\alpha} 
\left(1-\chi\sqrt{1+\varepsilon(\frac{a_i}{a_o})^4}\right) 
 \label{fdef} \nonumber 
\end{eqnarray}
The only formal difference between the Friedmann equation~(\ref{gfeq}) and       
the corresponding RS equation is that the {\em dark radiation} term~\cite{Ma}    
proportional to $a^{-4}_o$ does not appear explicitly.  It has been included     
in $\tilde{\Lambda}_\ast$. Actually, at zero order in $\alpha$ we have:          
$\tilde\Lambda_\ast = \tilde{\Lambda}_{RS} + 3Ca^{-4}_o + O(\alpha)\,,$          
and we recover the Friedmann equation corresponding to RS braneworlds            
(see, e.g.,~\cite{Friedmann}). Moreover, in analogy to the RS case,                    
$\tilde{\kappa}^2$ is the 4D gravitational coupling constant.                    

As in standard cosmology, the Friedmann equation~(\ref{gfeq})
together with the energy-momentum tensor conservation equations
[a consequence of the divergence-free character of the left-hand
side of~(\ref{fieldeq})],
\begin{eqnarray}
\dot{\rho} = -3(\rho+p)H \,, \nonumber
\end{eqnarray}
and a barotropic equation of state $p=p(\rho)$, describe completely
the cosmological dynamics on the brane.

The explicit form of the bulk solution has been presented in~\cite{GeSo},        
for the case in which the fifth dimension is assumed {\em static}                
($\dot{b}=0$).  This solution is a 5D black hole spacetime which in              
the limit $\alpha\rightarrow 0$ coincides with the 5D Schwarzschild-AdS          
black hole.  We have also found that the integration constant $C$                
is proportional to the black hole mass.

%%%%%%%%%%%%%%%%%%%%%%%%%%%%%%%%%%%%%%%%%%%%%%%%%%%%%%%%%%%%%%%%%%%%%%
%
%     SECTION V.
%
%%%%%%%%%%%%%%%%%%%%%%%%%%%%%%%%%%%%%%%%%%%%%%%%%%%%%%%%%%%%%%%%%%%%%%

\section{Cosmological dynamics}                                                  

Let us now analyze some questions about the behaviour of the FLRW models         
according to this theory.  For a general analysis see~\cite{GeSo}. Here, we             
will be interested in the high-energy regime, where our theory                   
deviates significantly from General Relativity\footnote{This part                 
is based on work in progress in collaboration with Roy Maartens.} (GR).               

That regimes occurs at very early times, where we would expect a large           
contribution from the bulk curvature, i.e. $\varepsilon(\frac{a_i}{a_o})^4
\gg 1$.  In this situation the Friedmann equation can be approximated by
\be\label{exp}
H^2\simeq \frac{\tilde{\kappa}^2}{3\varepsilon}\left[\rho(1+
\frac{\rho}{2\lambda})+\frac{\lambda}{2}\right](\frac{a_o}{a_i})^4
+\frac{\tilde k^2\lambda}{6}\frac{\chi}{1-\chi}\sqrt{\varepsilon} 
(\frac{a_i}{a_o})^2\,.
\ee
Following the general belief that the expansion of the Universe went
through an early stage of  positive acceleration, let us assume that
that the Hubble function, for small $a_o/a_i$, behaves as $H^2\sim (a_o/a_i)^{-n}$
($n\leq 2$). 
If we take $n=2$ we get zero acceleration (which corresponds to the 
behaviour of the Milne Universe).  On the other hand, if we take $n=0$, we 
get the scale factor of the de Sitter model, which grows exponentially in 
time.  From our approximation (\ref{exp}), we can see that in order to 
have a positive acceleration, the first term on the right-hand side should 
dominate at some stage, and should have an expansion index $n<2$. In general,      
this is not possible at very early times, where the second term with an            
expansion index $n=2$, would dominate.  According to this, we could expect         
that the Universe had an initial quasi-Milne phase driven by geometry and          
afterwards a positive acceleration driven by, for instance, a scalar field.        
Then, in these models, an inflationary stage would be shifted in time.             
Moreover, they also have the interesting property that the particle 
horizon will be logarithmic divergent, since the Milne Universe doesn't have 
any horizon. This fact could suggest an alternative mechanism to solve 
the horizon problem, even if we still require an accelerated era. 
 
Let us now look at the Big-Bang Nucleosynthesis (BBN) and supernova               
constraints for our model\footnote{This part is based on work in progress        
in collaboration with Luca Amendola and Pier Stefano Corasaniti.}. 
First, it is clear that in the limit $\varepsilon=0$ the cosmology will be 
the standard one.  A surprise arises when one wants to study the BBN 
constraint in the strong bulk-curvature regime. %C 
Indeed, assuming that $1/\lambda$ and $\varepsilon(\frac{a_i}{a_o})^4$            
were large enough, the Hubble function would behave as follows 
\begin{eqnarray} 
H^2\simeq \frac{\tilde k^2 a_o^4}{6\varepsilon\lambda a_i^4}\rho^2\,, \nonumber 
\end{eqnarray} 
then, since during the BBN the dominant energy-density component would be         
the radiation one, i.e. $\rho\propto a_o^{-4}$, this implies that 
$H^2\propto a_o^{-4}$, which means that we recover the same dynamical 
behaviour as in GR.  Using the standard BBN calculations \cite{BBN} it 
is possible to find a range of parameters compatible with the observations 
and the requirement that the quadratic part on the energy density 
be the dominant one.                         
 
\begin{figure}\label{fig} 
\centerline{\includegraphics[width=12pc]{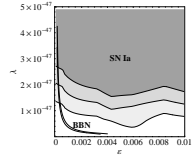}} 
\caption{BBN and SNIa constraints to large deviations of our model                 
from the GR one, for $\chi=0.2$ and $a_i=1$.
The coloured regions from the darker to the white correspond to 1 to 4 
$\sigma$ confidence form the SNIa data \cite{SN}. BBN and SNIa are in 
agreement in our model at 2 $\sigma$. $\lambda$ is measured in $GeV^4$.}
\end{figure}

The second observational constraint we are going to consider is the one           
imposed by supernovae (SNIa) redshift measurements \cite{SN}. If we combine the 
constraints coming from this data with the previous ones from BBN we get a 
tiny region where the cosmological constraints are compatible with a strong 
deviation from the GR behaviour, as it can be seen from Fig.(\ref{fig}).          
Even if the cosmological constraints are satisfied, the very small               
magnitude of $\lambda$ is quite suspicious.  Indeed in the RS case, 
from deviation of Newton law in sub-mm experiments,  
one gets the constraint $\lambda>10^8\ GeV^4$ \cite{New}.
In any case,
the implementation of such constraint in this case is still an open question.

%%%%%%%%%%%%%%%%%%%%%%%%  FIGURE  %%%%%%%%%%%%%%%%%%%%%%%%%%%%%%%%%%%%%%%%%%%%%%
 
%%%%%%%%%%%%%%%%%%%%%%%%%%%%%%%%%%%%%%%%%%%%%%%%%%%%%%%%%%%%%%%%%%%%%%%%%%%%%%%%

To sum up, we have studied the dynamics of the standard FLRW cosmological          
models in a theory that generalizes the RS model~\cite{RaSu}                      
by taking into account the Gauss-Bonnet higher-order curvature 
term~\cite{BoDe}.  We have shown how the equations governing the 
cosmological dynamics can be written in the same form as those in RS scenarios 
but with time-dependent 4D gravitational and cosmological constants.  Studying 
the 5D geometry of our model we have found that the time variation of 
the constants is parametrized only by the mass of a black hole in the 
bulk. We have also shown that the higher-order curvature terms present in 
our theory, which are dominant at high energies and change the cosmological 
dynamics at early times, can provide alternative cosmological 
scenarios for the study of unsolved cosmological problems. 
In this sense, we have seen that a strong bulk curvature limit is compatible        
with BBN \cite{BBN} and SNIa \cite{SN} observational data, and how this could 
provide alternative solutions to the horizon problem.  In this respect,                    
the compatibility of such a model with the constraints imposed by                  
the Newtonian gravity at low energies is still open.

%%%%%%%%%%%%%%%%%%%%%%%%%%%%%%%%%%%%%%%%%%%%%%%%%%%%%%%%%%%%%%%%%%%%%%
%
%     ACKNOWLEDGEMENTS
%
%%%%%%%%%%%%%%%%%%%%%%%%%%%%%%%%%%%%%%%%%%%%%%%%%%%%%%%%%%%%%%%%%%%%%%

\acknowledgements

C.G. is supported by a P.P.A.R.C. studentship.
C.F.S. is supported by the E.P.S.R.C. The authors wish to thank
Luca Amendola, Pier Stefano Corasaniti and Roy Maartens.

\end{article}

\end{document}